\documentclass[11pt]{article}

\usepackage[russian,english]{babel}
\usepackage[tbtags]{amsmath}
\usepackage{amsfonts,amssymb}

\usepackage{mathrsfs}
\usepackage[short]{mz2ewin}
\usepackage{graphicx}

\usepackage[pdfauthor={Brezhnev}, colorlinks=true, linktocpage=false,
            pdfwindowui=false, pdfmenubar=false, 
            citecolor=blue,
            hyperfootnotes=true, pdfstartview=FitH,
            bookmarksopen=false, bookmarks=false]{hyperref}

\newtheorem{theorem}{\textbf{\emph{Theorem}}}
\theoremstyle{definition}
\newtheorem{remark}{\textbf{\emph{Remark}}}

\def\bop{\boldsymbol{\frak p}}
\def\po{\frak p_\circ}

\begin{document}

\author{\scshape\mdseries Yu.\,V.\,Brezhnev}
\address{Tomsk State University\\
Tomsk, 634050 Russia}
\email{brezhnev@mail.ru}

\author{\scshape\mdseries A.\,V.\,Tsvetkova}
\address{
Ishlinsky Institute for Problems in Mechanics RAS\\
Moscow, 119526 Russia\\
Moscow Institute of Physics and Technology\\Dolgoprudnyi, Moscow
Region, 141701 Russia}

\email{annatsvetkova25@gmail.com}

\title{On Hamiltonian systems integrable in elliptic functions
that describe waves over underwater banks and ridges}
\maketitle
\thispagestyle{empty}

\parbox{9cm}{\footnotesize\emph{Abstract}. We discuss the
4-dimensional Hamiltonian systems that describe waves over
underwater banks and ridges. The systems are exactly integrable in
terms of elliptic functions and of solutions to nontrivial
transcendental equations involving the elliptic integrals
(Weierstrass' $\zeta$-function).}

\tableofcontents

\section*{\!\!Introduction}
In this paper, we discuss the problem of integration of the
following Hamiltonian system with two degrees of freedom:
\begin{equation}\label{Ham_syst}
\dot p=-H_x,\quad \dot x=H_p,\quad
H=|p|C(x),
\end{equation}
where $C(x)$ is a positive smooth function.

We show that such systems are integrable in elliptic functions for
functions $D(x)$ of a certain form; we shall write them out further
below. The problem of integration is interesting not only as an
independent problem, but also from the standpoint of applications.
In particular, such Hamiltonian systems arise in the study of
rapidly varying solutions of the two-dimensional wave equation. For
example, in the case of long linear gravity surface waves in an
infinite basin with varying bottom given by $D(x) > 0$, the function
$C(x)=\sqrt{g D(x)}$ ($g$ is the free-fall acceleration), and the
corresponding equation has the form \cite{Stoker,Peli}
\begin{equation}\label{Weq1}
\frac{\partial^2 u}{\partial t^2}-\langle \nabla,
C^2(x)\nabla\rangle u=0,\quad C^2(x)=gD(x),\quad x\in \mathbb{R}^2.
\end{equation}
The initial (or some other additional) conditions for system
\eqref{Ham_syst} are generated by the conditions for equation
\eqref{Weq1}. We consider the initial conditions for system
\eqref{Ham_syst} corresponding to the Cauchy problem for
\eqref{Weq1} with localized initial conditions
\begin{equation}\label{Weq2}
u|_{t={t_\circ}}=V\left(\frac{x-x^{\circ}}{l}\right),\quad
\left.\frac{\partial u}{\partial t}\right|_{t_{\circ}}=0,
\end{equation}
where $V (y)$ is a smooth function decreasing rapidly enough as
$|y|\to \infty$, the parameter $l$ characterizes the source size,
and $x^{\circ}$ is a point in whose neighborhood the initial
perturbation is localized. Such conditions for \eqref{Weq1} give the
following Cauchy problem \cite{DobSekTirVol,DobShafTir,DobrNaz} for
system \eqref{Ham_syst}:
\begin{equation}\label{cond}
p|_{t_\circ}=\mathbf{n}(\psi)=\begin{pmatrix} \cos\psi\\
\sin\psi\end{pmatrix}, \quad \psi\in[0,2\pi],\quad
x|_{t_\circ}=x^{\circ}\in \mathbb{R}^2.
\end{equation}
The Hamiltonian system we consider makes it possible to describe the
water waves.

By $X(\psi,t),P(\psi,t)$ we denote the solutions of problem
\eqref{Ham_syst}, \eqref{cond}. At each moment of time $t$, the ends
of these trajectories determine the following smooth closed curves
\begin{equation*}
\Gamma_t=\{x=X(\psi,t),\quad p=P(\psi,t),\quad \psi \in S^1\}
\end{equation*}
in the 4-D phase space $\mathbb R^4_{xp}$, which are called
\emph{wave fronts in the phase space}. According to
\cite{DobSekTirVol,DobShafTir,DobrNaz}, at each $t$, the asymptotic
solution of problem \eqref{Weq1}, \eqref{Weq2} is determined by
$\Gamma_t$ and turns out to be localized in a neighborhood of the
curves $\gamma_t=\{x=X(\psi,t),\;\psi \in S^{1}\}$---the projections
of $\Gamma_t$ on $\mathbb{R}_{xp}^4$---which are called \emph{fronts
in the configuration (physical) space}. In contrast to $\Gamma_t$,
the curves $\gamma_t$ may be non-smooth and have the
self-intersection points. Under $t=0$, the set $\gamma_t$ is the
point~$x^{\circ}$.

As mentioned above, we here we will show that, in the cases where
the bottom is shaped as an underwater bank, i.e., the depth of the
water layer is determined by the function
\begin{equation}\label{bank}
D(\varrho,\varphi)=D(\varrho)\equiv
\frac{b+\varrho^2}{a+\varrho^2},\,\qquad a>b>0\;\;\text{are constants},\quad
\varrho\;\text{\;is the polar radius},
\end{equation}
and in the case where the underwater ridge-shape is determined by
the function
\begin{equation}\label{ridge}
D(x_1,x_2) = \frac{x_1^2+b}{x_1^2+a},\qquad a>b>0\;\;\text{are constants}\,.
\end{equation}
The above-written Hamiltonian system is integrable in elliptic
functions. The exact analytic solutions obtained in these cases are
given in sect.~\ref{result}. These formulas can be used to construct
an asymptotic solution of problem \eqref{Weq1}, \eqref{Weq2} and
asymptotic solutions of stationary equations with localized
right-hand sides, for example, the Helmholtz equation \cite{Anikin}
$$
-\omega^2 v-\varepsilon^2\langle \nabla, C^2(x)\nabla\rangle
v=\frac{1}{\varepsilon}V\left(\frac{x-x^{0}}{\varepsilon}\right), \qquad
\omega=\mathrm{const},\quad x\in \mathbb{R}^2.
$$
The corresponding formulas will be given in the expanded version of
the work. In sect.~\ref{visual}, we describe an algorithm for
constructing fronts, which makes it possible to visualize them and
facilitates dealing with the obtained expressions. In
sect.~\ref{proof}, we outline the proofs.

\section{Exact analytic solutions}\label{result}

Since the function determining the bottom shape is symmetric in both
cases under study, we can without loss of generality assume that
$x^{\circ}= \left(\begin{smallmatrix}
-\xi\\0\end{smallmatrix}\right)$, where $\xi>0$.

Let us consider the situation where the function describing the
depth of the water layer has the form \eqref{bank}, i.e., determines
the underwater bank. We assume that the free-fall acceleration
$g=1$. Then, in the polar $(\varrho,\varphi)$-coordinates
$\{x_1=\varrho \cos\varphi,\,x_2=\varrho \sin\varphi\}$, the
Hamiltonian becomes
\begin{equation}\label{Ham_new} \mathscr
H(\varrho,\varphi;u,v)=\sqrt{u^2+\frac{v^2}{\varrho^2}}\cdot
\sqrt{\frac{\varrho^2+b}{\varrho^2+a}},
\end{equation}
where $u:=p_{\varrho}$, $v:=p_{\varphi}$ are the momenta
corresponding to the system $(\varrho,\varphi).$ The initial
conditions \eqref{cond} in these coordinates become
\begin{equation}\label{InD2}
u|_{t_\circ}=-\cos \psi,\qquad v|_{t_\circ}=-\xi\sin
\psi,\qquad \varrho|_{t_\circ}=\xi,\qquad \varphi|_{t_\circ}=\pi.
\end{equation}

\begin{theorem}\label{theorem1}
The solution of the Hamiltonian system with Hamiltonian
\eqref{Ham_new} and initial conditions \eqref{InD2} is given by the
expressions
\begin{equation}\label{SOL}
\left\{
\begin{aligned}
\varrho&=\sqrt{\wp(\bop)+\delta-a},\\
 \varphi&=\pi \pm \bigg\{1+2\,b\,
\frac{\zeta(\varkappa)}{\wp'(\varkappa)}\bigg\}h\cdot(\bop-\po) \pm
\frac{b\,h}{\wp'(\varkappa)}
\ln\frac{\sigma(\bop-\varkappa)\,\sigma(\po+\varkappa)}
{\sigma(\bop+\varkappa)\,\sigma(\po-\varkappa)},\\
u&=-\frac12
\sqrt{\frac{\xi^2+b}{\xi^2+a}}\,
\frac{\wp'(\bop)}{\wp(\bop)+\delta-a+b}\,
\frac{1}{\sqrt{\wp(\bop)+\delta-a}},\qquad
v=-\xi\sin\psi\,,
\end{aligned}
\right.
\end{equation}
where the variables $\bop, \po$ and $\varkappa$ are solutions of the
following transcendental equations
\begin{equation}\label{p0pkappa}
\wp(\po)=
\xi^2-\delta+a, \quad \wp(\varkappa)=\frac13(a+b-\frak h),\qquad
t-t_\circ=\delta\cdot
(\bop-\po)-\zeta(\bop)+\zeta(\po).
\end{equation}
The expressions for the constants $\delta, \alpha$, and $\beta$ in
terms of the parameters of the problem have the form
\begin{equation}\label{alpha}
\begin{aligned}
\delta&=\frac{1}{3}\,({\frak h}+2\,a-b), \qquad
\alpha=\frac{4}{3}\,\Bigl({\frak h}^2-
2\,(a-2\,b)\,{\frak h}+a^2-a\,b+b^2\Bigr),\\[1ex]
\beta&=\frac {4}{27}\,\Bigl({\frak h}-a+2\,b\Bigr)
\Bigl(2\,{\frak h}^2-4\,(a-2\,b){\frak h}+(2\,a-b)(a+b)\Bigr),
\end{aligned}
\end{equation}
where
\begin{equation}\label{h}
h^2={\frak h}=\xi^2\,\frac{\xi^2+a}{\xi^2+b}\sin^2\psi\,.
\end{equation}
Here, as always in the sequel, the notation
$\wp(z)=\wp(z;\alpha,\beta)$, $\zeta(z)=\zeta(z;\alpha,\beta)$,
$\sigma(z)=\sigma(z;\alpha,\beta)$ is used for the Weierstrass
elliptic functions \cite{Ahiezer}.
\end{theorem}
\begin{remark}
Note that $\po, \bop$, and $\varkappa$ need not be real. They lie on
the edges of the parallelogram $(0,\omega,\omega+\omega',\omega')$,
where $\omega$ is the pure real period of the Weierstrass
$\wp$-function and $\omega$ is the pure imaginary one. In this case,
$\bop$ lies on the same edge of the parallelogram as $\po$. We also
note that equations \eqref{p0pkappa} have infinitely many roots. For
definiteness, we assume that the roots are taken from the first
positive branch. In sect.~\ref{visual}, we describe an algorithm for
constructing fronts, which illustrates these formulas.
\end{remark}

In the case where the function determining the depth of the water
layer has the form \eqref{ridge} and describes the underwater ridge,
the solutions are given by the following theorem.
\begin{theorem}
The solution of the Hamiltonian system \eqref{Ham_syst} with
Hamiltonian
$$
\mathscr H(x,y;u,v)=
\sqrt{p_{x_1}^2+p_{x_2}^2}\sqrt{(x_1^2+b)/(x_1^2+a)}
$$
and initial conditions \eqref{cond} has the form
\begin{equation}\label{res_case2}
\left\{
\begin{aligned}
x_1&=-\sqrt{\wp(\bop;\alpha,\beta)+\delta-a},\qquad
x_2=\sqrt{1-\frak h^{-1}}t+ \sqrt{\frak
h-1}\,(b-a)\cdot(\bop-\po),\\
p_{x_1}&=\frac{1}{2}\,\frac{\gamma}{\sqrt{\frak h-1}}
\frac{\wp'\big({\frak
p}(t)\big)} {\wp\big({\frak p}(t)\big)+\delta-a+b}
\sqrt[-2]{\wp\big({\frak p}(t)\big)+\delta-a},\qquad
p_{x_2}=\gamma.
\end{aligned}
\right.
\end{equation}
The variables $\bop$ and $\po$ are solutions of the transcendental
equations
\begin{equation}\label{new2'}
\frac{1}{h}\,t=\delta\cdot
(\bop-\po)-\zeta(\bop;\alpha,\beta)+
\zeta(\po;\alpha,\beta),\qquad
\po=\wp^{-1}(\xi^2-\delta+a;\alpha,\beta).
\end{equation}
The expressions for the constants $\delta, \alpha$, and $\beta$ in
terms of the parameters of the problem have the form
\begin{equation}\label{alpha'}
\begin{aligned}
\delta&=\frac{1}{3}\,\Bigl((b-a)\,
\frak h+3\,a-2\,b\Bigr), \qquad
\alpha=\frac{4}{3}\,\Bigl((b-a)^2\frak h^2-
(b-a)\,b\,\frak h+b^2\Bigr),\\[1ex]
\beta&=
\frac {4}{27}\,\Bigl((b-a)\,\frak h+b\Bigr)
\Bigl((b-a)\,\frak h-2\,b\Bigr)\Bigl(2\,(b-a)\,\frak h-b\Bigr),
\end{aligned}
\end{equation}
where
\begin{equation}
h^2=\frak{h}=\frac{\xi^2+b}{(\xi^2+b)-(\xi^2+a)\sin^2\psi}, \qquad
\gamma=\sin \psi.
\end{equation}
\end{theorem}

\section{Visualizing the fronts. An algorithm}\label{visual}

Here we describe an algorithm for constructing, in particular, the
fronts $\gamma_t$ by using the software package
\textsc{Wolfram~Mathematica}. This algorithm makes it easy to visualize
the formulas obtained in the preceding section and illustrate the
application of these formulas. On the other hand, the problem of
visualizing fronts is also interesting from the standpoint of
applications, because, as mentioned above, the asymptotic solution
of problem \eqref{Weq1}, \eqref{Weq2} is localized in a neighborhood
of the front \cite{DobSekTirVol,DobShafTir,DobrNaz}.
\begin{figure}[h!] \centering
\includegraphics[width=0.7\linewidth]{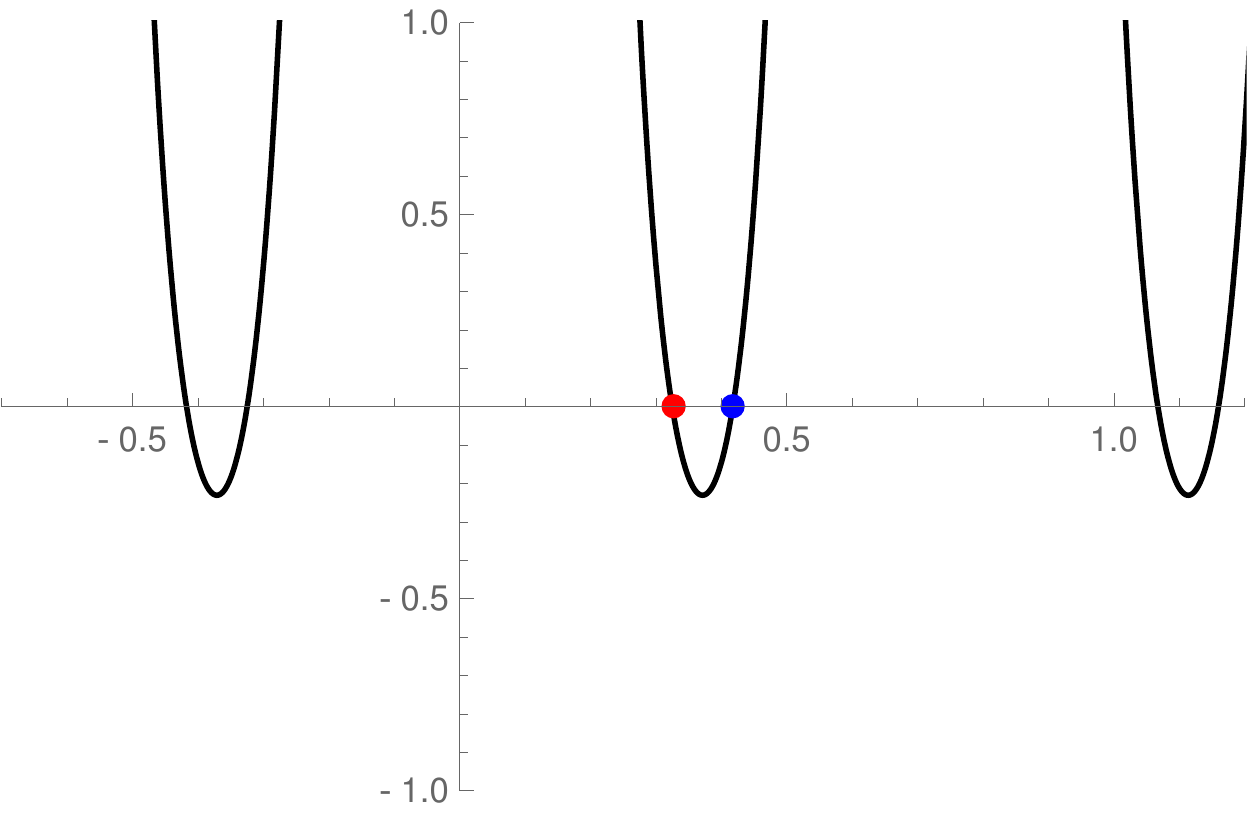}
\caption{The graph of $y=\wp(x)-\xi^2+\delta-a$ under
$\psi=\frac{9}{10}\pi$}\label{figp0}
\end{figure}
\begin{figure}[h!]
\centering
\begin{minipage}[h]{0.7\linewidth}
\begin{minipage}[h]{0.6\linewidth}
{\includegraphics[width=0.85\linewidth]{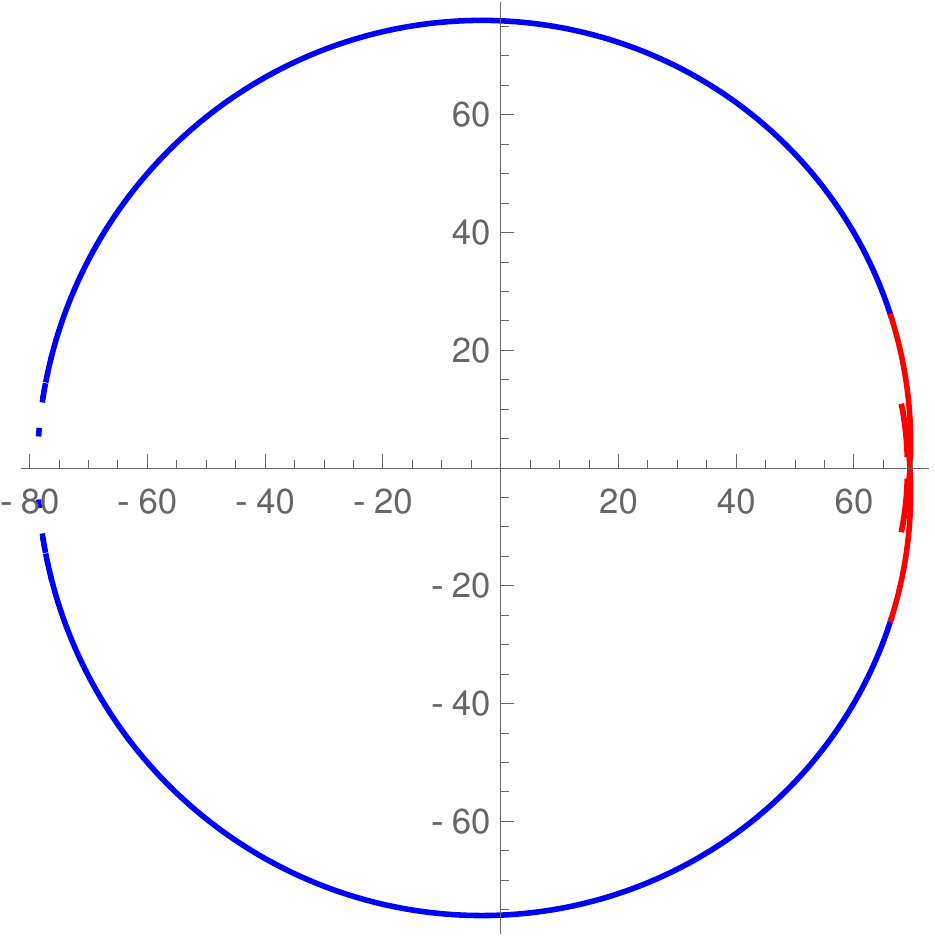}}
\end{minipage}
\hfill
\begin{minipage}[h]{0.35\linewidth}
{\includegraphics[width=0.6\linewidth]{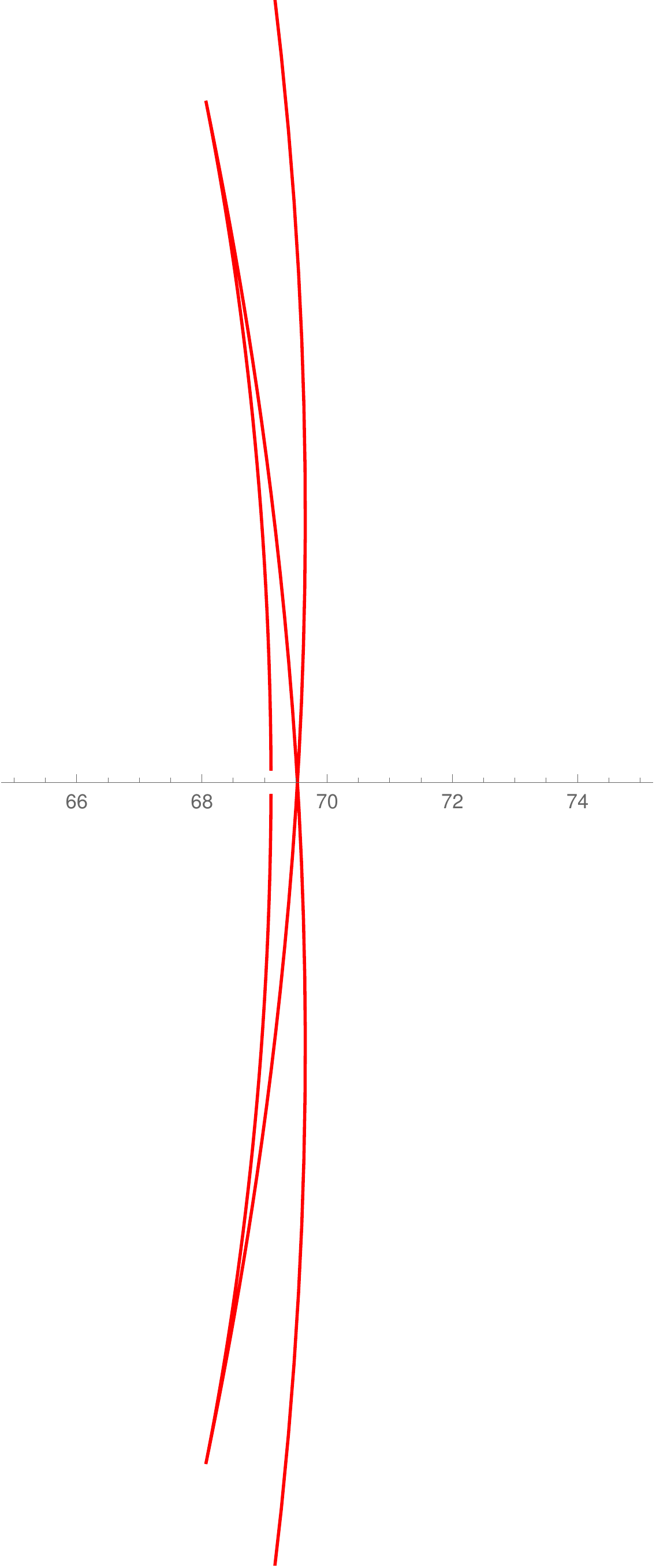}}
\end{minipage}
\caption{The graph of the front at $t=100$} \label{figt100}
\end{minipage}
\end{figure}

\section*{\!\!Algorithm}

\emph{Step}-1: We fix the following parameters of the problem: the
constants $a=100$, $b=1$ determining the bottom shape and the
constant $\xi=0.5$ characterizing the source position.

\emph{Step}-2: For each fixed $\psi \in [0,\pi]$, there exist
infinitely many roots of the equation
\begin{equation}\label{alg_p2}
\wp(p_{\circ})=\xi^2-\delta+a.
\end{equation}
As $p_{\circ1}(\psi)$ and $p_{\circ2}(\psi)$ we take, respectively,
the first and second positive roots (see Fig.~\ref{figp0}). In the
figure, the red point corresponds to $p_{\circ1}$ and the blue
point, to $p_{\circ2}$. We determine $\varkappa(\psi)$ by the
formula \eqref{p0pkappa}, where we again take the first positive
root.
\begin{remark}
As already mentioned, the roots of \eqref{alg_p2} need not be real.
We must seek the complex roots in one of the forms $ip_{\circ}$,
$p_{\circ}+\omega'$, and $ip_{\circ}+\omega$, where $\omega$ is the
pure real period of the Weierstrass $\wp$ and $\omega'$ is its pure
imaginary period, $p_{\circ}$ is real; as  $p_{\circ1}$ and
$p_{\circ2}$ we must again take the first two positive roots.
\end{remark}

\emph{Step}-3: We fix a moment of time $t$. Let
$p_{i}(\psi,t)\,(i=1,2)$ be the first positive root of the equation
$t=\delta\cdot [p-p_{\circ i}(\psi))-\zeta(p)+\zeta(p_{\circ
i}(\psi)]$. If $p_{\circ i}$ is not real, then we seek $p_{i}$ in
the same manner as $p_{\circ i}$.

\emph{Step}-4: For each fixed $t$, we plot the curves in polar
coordinates $\varrho,\varphi$, using formulas \eqref{SOL}. In the
case $\psi \in \left[0,\frac{\pi}{2}\right]$, we take $p_{\circ1}$
and $p_1$ as parameters, and in the case $\psi \in
\left[\frac{\pi}{2},\pi\right]$, we take $p_{\circ2}$ and $p_2$.

\emph{Step}-5: By symmetry, we reflect the obtained graph about the
horizontal axis. Figure~\ref{figt100} shows the front at $t = 100$;
the dashed curve in this figure corresponds to $\psi \in
\left[0,\frac{\pi}{2}\right]$ and the solid curve, to $\psi \in
\left[\frac{\pi}{2},\pi\right]$.

\section{Proof of theorem \ref{theorem1} in a nutshell}\label{proof}

Solving the Hamiltonian system \eqref{Ham_syst} with Hamiltonian
\eqref{Ham_new} and initial conditions \eqref{InD2}, we get
\begin{equation*}
v=\gamma \quad (=\mathrm{const})\,,\quad
\left(u^2+\frac{\gamma^2}{\varrho^2}\right)
\frac{\varrho^2+b}{\varrho^2+a}=\frac{\gamma^2}{h^2}
\quad( = \mathrm{const}).
\end{equation*}
Making use of the integral $\mathscr
H(\varrho,\varphi;u,v)=\gamma/h$, we obtain the following autonomous
dynamics for the variable $\varrho$
\begin{equation*}
\varrho\,\dot\varrho=
\frac{\sqrt{(\varrho^2+b)\left[\varrho^2(\varrho^2+a)-
(\varrho^2+b)h^2\right]}}{(\varrho^2+a)}\,.
\end{equation*}
Making the change $z=\varrho^2+a$ and applying the shift
$s=z-\delta$ to reduce the integral to the canonical Weierstrass
form
\begin{equation*}
\int\limits^{\varrho^2+a-\delta}_\infty\!\!\!
\frac{(s+\delta)\,ds}{\sqrt{4\,s^3-\alpha\,s-\beta}}=t,
\end{equation*}
we derive an expression for $\varrho$ in system \eqref{SOL}. Using
the obtained result and the integral $\mathscr{H}=\gamma/h$, we
obtain the formula for $u$. Substituting the obtained expression
into the Hamiltonian system, one obtains the equation
\begin{equation*}
\dot \varphi=
h\left(1+ \frac{b}{\wp(\bop)+\delta-a}
\right)\frac{1}{\wp(\bop)+\delta}\,.
\end{equation*}
Its integration leads to the logarithmic elliptic integral that can
be easily evaluated \cite{Ahiezer}. Simplifying the obtained
expression, we arrive at the solutions \eqref{SOL}.

\section*{\!\!Acknowledgments}

The authors wish to express their gratitude to S.~Yu.~Dobrokhotov
for the statement of the problem and to M.~Babich and M.~Pavlov for
useful discussions.


\end{document}